\documentclass[preprint,prb,showpacs,superscriptaddress,showkeys]{revtex4}
\usepackage{amsfonts}
\usepackage{amsmath}
\usepackage{amssymb}
\usepackage{graphicx}
\usepackage{subfigure}

\begin{document}

\title{Interplay of defect cluster and the stability of xenon in uranium dioxide by density functional calculations}
\author{Hua Y. Geng}
\affiliation{National Key Laboratory of Shock Wave and Detonation
Physics, Institute of Fluid Physics, CAEP;
P.O.Box 919-102 Mianyang, Sichuan, P. R. China, 621900}
\author{Ying Chen}
\affiliation{Department of Systems Innovation, The University of Tokyo, Hongo 7-3-1, Tokyo 113-8656,
Japan}
\author{Yasunori Kaneta}
\affiliation{Department of Systems Innovation, The University of Tokyo, Hongo 7-3-1, Tokyo 113-8656,
Japan}
\author{Motoyasu Kinoshita}
\affiliation{Nuclear Technology Research Laboratory, Central Research Institute of Electric Power Industry, Tokyo 201-8511, Japan}
\affiliation{Japan Atomic Energy Agency, Ibaraki 319-1195, Japan}
\author{Q. Wu}
\affiliation{National Key Laboratory of Shock Wave and Detonation
Physics, Institute of Fluid Physics, CAEP;
P.O.Box 919-102 Mianyang, Sichuan, P. R. China, 621900}

\keywords{fission gas, defect cluster, nonstoichiometric oxides, uranium dioxide, fluorite structure}
\pacs{61.72.J-, 71.15.Nc, 71.27.+a}

\begin{abstract}
Self-defect clusters in bulk matrix might affect the thermodynamic behavior of fission gases
in nuclear fuel such as uranium dioxide.
With first-principles LSDA+\emph{U} calculations and taking xenon as a prototype,
we find that the influence of oxygen defect clusters on the thermodynamics of gas atoms
is prominent, which increases the solution energy of xenon by a magnitude of 0.5\,eV,
about 43\% of the energy difference between the two lowest lying states at 700\,K.
Calculation also reveals a thermodynamic competition between the uranium vacancy and tri-vacancy
sites to incorporate xenon in hyper-stoichiometric regime at high temperatures.
The results show that in hypo-stoichiometric regime
neutral tri-vacancy sites are the most favored position for diluted xenon gas, whereas in hyper-stoichiometric condition
they prefer to uranium vacancies even after taking oxygen self-defect clusters into account
at low temperatures, which not only confirms previous studies but also extends the conclusion to more realistic fuel operating conditions.
The observation that gas atoms are ionized to a charge state of Xe$^{+}$ when at a uranium vacancy site
due to strong Madelung potential implies that
one can control temperature to tune the preferred site of gas atoms and then the bubble growth rate.
A solution
to the notorious meta-stable states difficulty that frequently encountered
in DFT+$U$ applications, namely, the quasi-annealing procedure, is also discussed.

\end{abstract}

\volumeyear{year}
\volumenumber{number}
\issuenumber{number}
\eid{identifier}
\maketitle

\section{INTRODUCTION}
\label{sec:intr}

The thermodynamics of fission products in uranium dioxide has
been a focus of considerable experimental and theoretical attentions
in nuclear industry.
Xenon as the most important fission gas is one of them.
Concern has been particularly centered on xenon's
role in fuel \emph{swelling}---that could increase the pressure on the cladding of the fuel
rod under irradiation
and lead to rupture. A similar risk also exists for the
container of nuclear waste
in storage conditions.
This has accordingly led to a desire
to obtain a greater
understanding of the basic processes governing the migration
and trapping of xenon within
the fuels.\cite{grimes91,jackson85,bubble,nicoll95,noirot08}

Previous theoretical studies on xenon behavior employed interatomic potentials such as
shell model.\cite{grimes91,jackson85,nicoll95} This method provided
qualitative understanding of gas properties.
However, since shell model has severe transferability difficulty,\cite{geng07b}
the reliability of its results requires further verification by other methods.
Application of quantum mechanics to this problem
was available only recently and focused mainly on single gas atoms that occupying
point vacancies and Schottky sites.\cite{crocombette02,iwasawa06,freyss06,nerikar09a,nerikar09,yu09} In uranium dioxide, however, oxygen defect clusters dominate
and the interplay of them and fission gases might be the key to correctly understand the subtle material behavior.
For example, in hyper-stoichiometric regime of UO$_{2+x}$ where $x>0$,
oxygen self-defect cluster---the cub-octahedron (COT) cluster dominates when temperature is relatively low.\cite{geng08b,bevan86,cooper04,garrido03,nowicki00}
There is a big cavity at the center of COT, which can either be
empty (denoted as COT-v), or be filled by additional oxygen and forms COT-o cluster,
or be filled by xenon atom and becomes COT-xe. Furthermore, existence of COT-v and COT-o clusters
changes the concentrations of all other defect traps that the gas atoms can incorporate with.
Such kind of direct and indirect effects of oxygen clusters have not yet been investigated.
Big cavity also can be found at uranium vacancy or tri-vacancy (tri-V, a kind of bound Schottky).
We will show that xenon atoms are prone to occupying these traps
and become xenon-trap aggregates.
This incorporation behavior not only reduces elastic strains that imposed on
the bulk matrix but also changes the development of intra-granular bubbles, and thus is possible to alleviate
the fuel swelling that suffered from fission gases.

Energetics of xenon in defective UO$_{2}$ was modeled in
a $2\times2\times2$ supercell consisting of eight fluorite cubic unit cells.
Periodic boundary conditions and the density functional theory in local
spin density approximation with Hubbard
correction to the on-site coulombic repulsion
of the localized uranium 5$f$ orbitals (LSDA+\emph{U})
were employed to compute the total energy.\cite{kresse96,anisimov91,anisimov93} All structures
were fully relaxed until residual forces less than 0.01\,eV/\AA.
Details of the computational setup and the validation of the method are referred to
Refs.[\onlinecite{geng08,geng08b,geng08c}].
Particularly the LSDA+\emph{U} approach has been applied to
perfect\cite{geng07} and self-defective\cite{geng08,geng08b,geng08c}
UO$_{2}$ successfully, and yielded results in good agreement with experiments.
Oxygen defect clustering
and the relevant thermodynamics have also been well described by this
method.\cite{geng08b,geng08c}

In next section we will discuss a solution to the notorious
meta-stable states problem that frequently encountered in DFT+\emph{U}
applications. This approach was developed in our previous calculations.\cite{geng08,geng08b,geng08c}
Though it lacks a rigorous theoretical basis and
cannot guarantee that the true ground state can always be achieved,
we found that it is effective to reduce the frequency of encountering high-lying meta-stable states and
thus improves the reliability of the computational results.
In Sec. \ref{sec:EF} and \ref{sec:ES} a systematic analysis of the electronic structure
and energetics of xenon atoms that incorporating in nuclear fuels will
be given, as well as the influence of oxygen defect clustering on
the incorporation energies, the solution energies and the relevant thermodynamics of xenon gas.
In section \ref{sec:comp} a comparison with other theoretical results will be discussed, followed by a summary.

\section{QUASI-ANNEALING PROCEDURE}
\subsection{Theoretical argument}

DFT+\emph{U} formalism improves the performance of density functional theory on strongly \emph{correlated} electronic
system by including a Hubbard correction to the on-site coulombic repulsion in a semi-empirical manner. The cost is, however,
introduced a lot of \emph{local minima} on the energy surface that obstructing energy minimization process,
and making electronic
optimization algorithms get stuck in meta-stable states. Monitoring
the occupation matrices (MOM) of the localized $f$ orbitals
can solve this problem partially by varying their initial values to search for the lowest state.\cite{larson07,dorado09,jomard08} But this option is not generally available in most DFT packages.
Also the approach is a try and error method, which cannot ensure that the global minimum has already been obtained
before all possible occupation matrices have been tried.
Another concern about MOM is the computational cost.
If spin degree is not considered, there are $C^{n}_{7}$ different ways of filling $n$ electrons in seven $f$ levels diagonally.
The number of different nondiagonal occupation matrices is much larger, but we can
reduce it to several times of the number of the diagonal case by assuming that all other occupations are insignificant.
Thus for each atom there are $mC^{n}_{7}$ different ways to fill the $f$ levels, with $m$ less than ten.
It is also the total number of the runs that are required
for each calculation if the simulation cell contains only one symmetry inequivalent atom.
Unfortunately the symmetry of defective system is usually low,
and has several non-equivalent atoms (say, $k$) with localized $f$ electrons.
In such a case a total number of
$(mC^{n}_{7})^{k}$ runs are necessary for each calculation. For UO$_{2}$, $n$ equals 2
and $m$ can take 3, thus gives about 60 different occupation matrices for each uranium.\cite{dorado09}
If point defects are concerned, there are at least two non-equivalent uranium atoms, and
the number of the total runs would increase
to 3600. For defect clusters, $k$ should be greater than 3 and it requires
millions of runs for each calculation to get the final result. This is a huge burden even for modern
supercomputers.

In a classical system, the meta-stable states difficulty can be tackled satisfactorily with \emph{annealing} procedure. Namely,
to remove the thermal kinetic energy of a system gradually and slowly so that all low-lying states have been visited
before picking out the ground solution. Similar concept can be applied to electronic system.
The basic idea is to shake or heat the electronic system with a spurious energy noise
to help it overcome the
energy barriers.
We call this method the \emph{quasi-annealing} (QA) procedure.

\begin{figure}
  \includegraphics*[width=2.5 in]{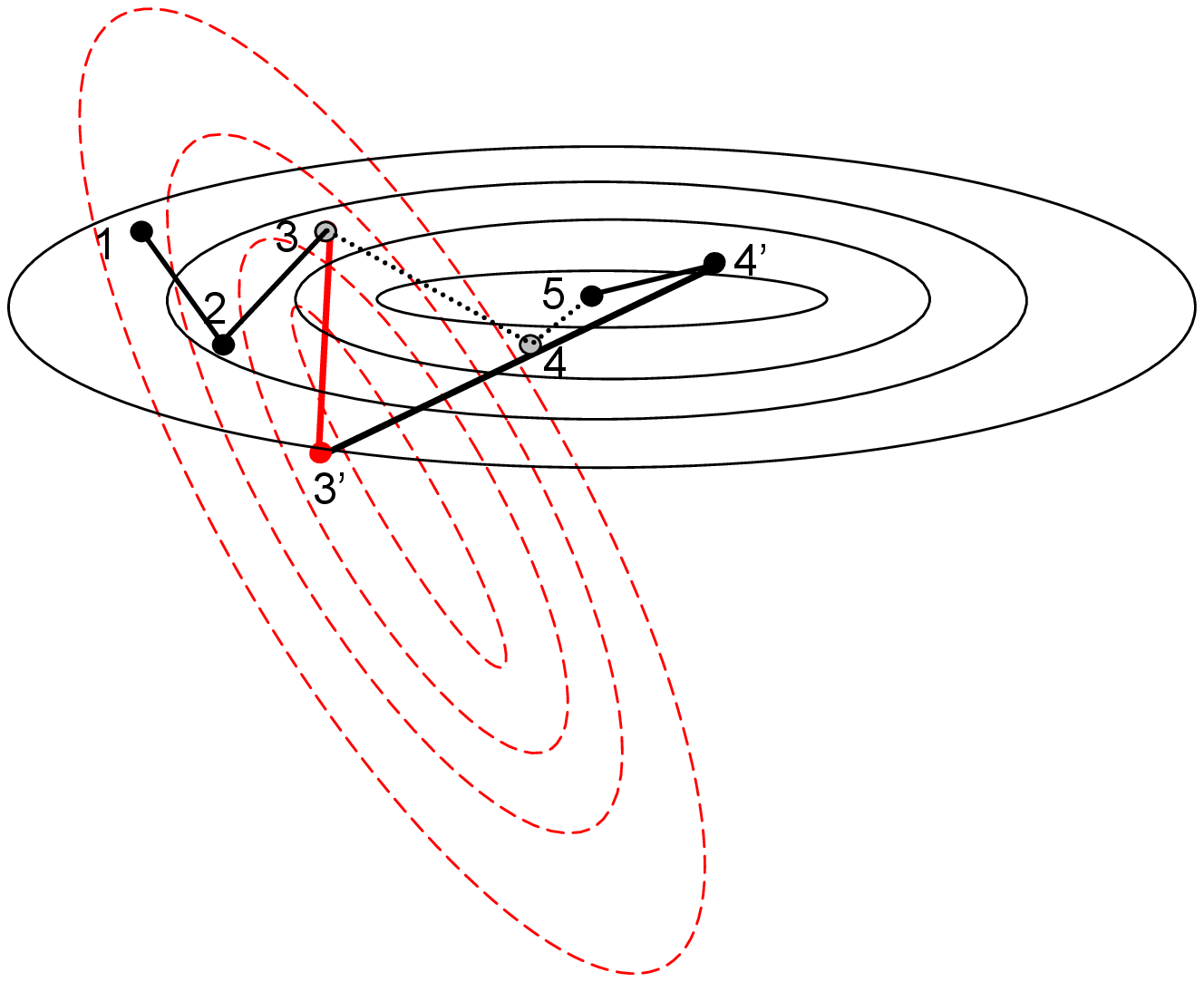}
  \caption{Schematic illustration of the electronic optimization process in
  quasi-annealing procedure, where fluctuations of potential surface
  (here from the solid to the dashed, and then back to the original solid contours)
  due to ionic drifts drag the minimizing path from 1, 2, 3, 4, 5
  to 1, 2, 3, 3$^{'}$, 4$^{'}$, 5, thus circumvents possible meta-stable states on the original route.
  }
  \label{fig:qa-draft}
\end{figure}

The theoretical basis is that the electronic energy is a functional of the electron density $n(r)$,
which is in turn a unique functional of the external
potential $v(r)$. One can then convey the spurious noise from the ionic subsystem
to the electronic
subsystem via $v(r)$. It amounts to $\int \Delta v(r)n(r)dr$, where $\Delta v(r)$ is the fluctuation of the external potential.
By switching off this spurious energy gradually, one can extract the
ground state in a similar way as its classical counterpart.
Alternatively, we can understand the mechanism of removal of the meta-stable states in QA procedure by tracking the electronic
minimization process:
the potential fluctuations
alter the minimization path iteratively, thus being capable of avoiding any possible meta-stable states
that lying on its route, as illustrated in figure \ref{fig:qa-draft}.

In practice, one might exploit the \emph{residual
energy} of the minimization process. The energy uncertainty $\delta E$ due to non-convergence
of the self-consistency field (SCF) gives a quasi-random fluctuation in ionic forces,
which in turn leads to a gaussian distribution of the ions with respect to their physical positions.
The potential fluctuation $\Delta v$ arising from this ionic drift
eventually \emph{heats} the electronic system up.
Generation of $\Delta v$ from the forces can be done with standard structure optimization algorithms.
That is, one iteratively relaxes the ionic structure with an electronic state having a SCF tolerance
of $\delta E$.
In this realization, the only one parameter---the residual energy in SCF---controls the
spurious noise in the
electronic system. Its value should be large enough at the beginning so that
the electronic system can travel freely in the phase space.
By decreasing $\delta E$
gradually, one converges the electronic system down to the ground state.

The merits of QA are not just that it can be used to tackle the meta-stable states. By coupling with ionic
relaxation, one can optimize the electronic and ionic states simultaneously.
It reduces the total computational
cost dramatically when structure optimization is also required, especially when DFT+\emph{U} formalism
is employed where the SCF convergence is very slow.
To avoid ions drifting too far away from the target
configuration, one needs to restore the structure after some ionic steps.
Usually allowing cell volume and
shape to vary improves the performance, because it not only extends the searching space,
but also shakes the system globally and breaks the \emph{symmetry} imposed by the Bravais lattice,
which is one of the main reasons that lead to
high-lying meta-stable states in strongly correlated system.\cite{larson07,dorado09}

\subsection{Validation}

\begin{figure}
  \includegraphics*[width=3.0 in]{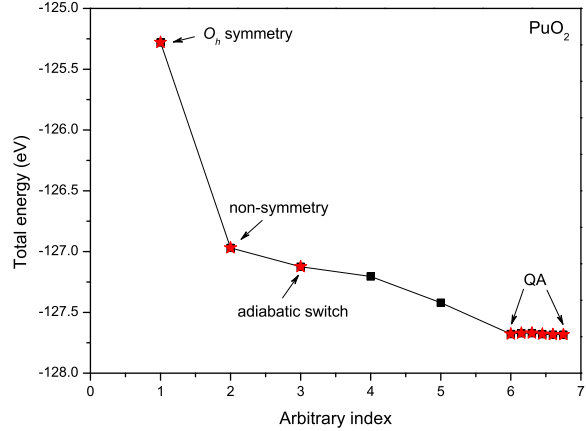}
  \caption{(Color online) Main meta-stable states in PuO$_{2}$ and the performance of quasi-annealing
  procedure, which predicted the lowest total energy. }
  \label{fig:qa-puo2}
\end{figure}

The QA procedure is summed up as follows, where in each step the computation restarts from the wave-function that was generated in its previous step:\cite{qanote}
\begin{enumerate}
  \item  Switch off symmetry, set appropriate values for SCF tolerance $\delta E$ and ionic relaxation step size
  $\delta r$.
  \item Employ standard ionic optimization algorithms to evolve the structure.
  \item Reduce $\delta E$ and $\delta r$ slightly, restore the structure,
  goto step 2 and repeat the procedure until $\delta E$ reaches the target precision.
  \item Conduct a standard SCF iteration.
  \item Slightly distort the structure, goto step 2 and repeat the whole process until no lower state can be found.
\end{enumerate}

At first we discuss the performance of QA in a perfect fluorite cubic cell of PuO$_{2}$ (with 12 atoms) that ordering in 1\textbf{k} antiferromagnetic
configuration with GGA+\emph{U} method.
The PAW pseudo-potentials and PBE exchange-correlation functional were used. The Hubbard parameters
were 4.0\,eV for $U$ and 0.7\,eV for $J$. A cutoff of 400\,eV was adopted for the kinetic energy
of the plane wave basis, and 63 irreducible \emph{k}-points were used to sample the Brillouin zone.
The lattice parameter was fixed at 5.45\,\AA. It should be pointed out that this setup was only for PuO$_{2}$.
All the following defects calculations were conducted with LSDA+\emph{U} method as detailed
in Sec.\ref{sec:intr}. Here we chose PuO$_{2}$ because this system has very stable meta-stable
states. For example if the cubic symmetry ($O_{h}$) is imposed onto the system,
one always obtains a meta-stable state having a total
energy of $-125.282$\,eV, no matter what the initial condition of the calculation is.
This is different from UO$_{2}$ where the symmetry-induced meta-stable state can be removed easily.
In PuO$_{2}$, however, one has to switch off the symmetry in order to get rid of this state.
As shown in figure \ref{fig:qa-puo2}, switching off the symmetry lowered the energy by 1.69\,eV. But this
is far from being the ground solution. Adiabatically switching on the Hubbard on-site interactions,
namely, increasing the $U$ and $J$ parameters from zero slowly, further
reduced the total energy about 0.16\,eV. The QA procedure, in contrast, predicted a much lower
energy. A total number of six independent QA runs were performed. The results were similar and the standard deviation $\sigma$ (scattering of the data) was 0.006\,eV.
On the other hand, direct SCF calculations have a standard deviation of two orders larger than QA.
This suggests that the electronic system has gotten rid of high-lying states and
converged closely to the ground solution in QA. The best result we ever had is $-127.684$\,eV.

It is helpful to compare QA results directly with MOM. We thus performed a GGA+\emph{U} calculation
on perfect UO$_{2}$ with the same setup as in Ref.[\onlinecite{dorado10}].
The only difference is that we used a 500\,eV cut-off energy for the plane-wave basis set and a 5$\times$5$\times$5
\emph{k}-point mesh instead of the 600\,eV cut-off and the 6$\times$6$\times$6 \emph{k}-point mesh that were used in that work.
This difference should have little influence on the final result since the total energy has already been converged
well with this setting of parameters. After a fully relaxation of the ionic structure, QA gave a total
energy of $-117.095$\,eV, lower than MOM's $-116.505$\,eV for fluorite structure and $-116.712$\,eV
for Jahn-Teller distorted geometry (see table \uppercase\expandafter{\romannumeral 5} in Ref.[\onlinecite{dorado10}]).
We cannot state that QA outperforms MOM, but it is obvious that an incomplete implementation of MOM
as done in Ref.[\onlinecite{dorado10}] does not necessarily lead to a ground state solution.

\begin{figure}
  \includegraphics*[width=3.0 in]{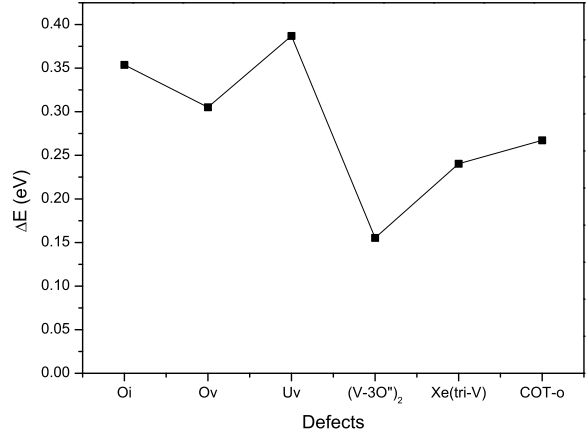}
  \caption{Total energy difference between direct SCF and QA calculations of
  some type of defects in UO$_{2}$, where the QA results are at the zero level. All direct SCF calculations
  were performed using the fully relaxed structures.}
  \label{fig:quasi-ann}
\end{figure}

Figure \ref{fig:quasi-ann} demonstrates the improvement of QA procedure against direct SCF calculations
for a set of defects in UO$_{2}$, including point oxygen interstitial, oxygen vacancy, uranium vacancy, split
quad-interstitial, xenon atom that incorporated in a tri-vacancy site, and COT-o cluster.
We found that direct SCF calculations always stopped at high energy states,
with an energy distance about 0.3\,eV to the QA results, regardless of the defect type.
It is worthwhile to point out that these direct SCF calculations were performed on structures that already optimized
by QA procedure, and the cubic symmetry of the lattice had been switched off.
One interesting example is about the stability of oxygen defect clusters.
Andersson \emph{et al.} did LSDA+\emph{U} calculations on the stability of defect clusters in UO$_{2}$ without
QA treatment.\cite{andersson09}
The results were controversial: they predicted that the split quad-interstitial ($V$-3O$^{''}$)$_{2}$ had an energy lower than COT-o
cluster.
Here we denote the split quad-interstitial by the symbol ($V$-3O$^{''}$)$_{2}$ because
it is actually a linear combination of two basic clusters---$V$-3O$^{''}$,
an oxygen vacancy surrounded by three Willis O$^{''}$ interstitials.\cite{geng08c}
A careful re-computation of the energetics of these two clusters with QA procedure, however,
gave a different picture. We found that Andersson \emph{et al.}'s energy of ($V$-3O$^{''}$)$_{2}$ was
almost the same as QA. But their value of COT-o was
much higher than that of QA (in terms of formation energy with respect to the same reference state, $-10.38$\,eV \emph{vs.} $-12.41$\,eV).
A direct SCF run with QA-optimized structure led to a meta-stable state with an energy
of 0.16\,eV higher than that of QA, and was much lower than Andersson \emph{et al.}'s
original result.
A re-check calculation using the same setup as theirs also failed to reproduce their results. According to the
variational principle of energy, we guess that their calculation of COT-o cluster might have stopped
at a meta-stable state. Unfortunately we cannot identify it without reproducing their electronic
state successfully.

Now we have two approaches that can tackle meta-stable states, \emph{i.e.}, MOM and QA.
In the case of perfect UO$_{2}$, though primary MOM calculations did not predict a lower state than QA,
it does not mean QA truly achieved the ground state.
We know that annealing procedure does not perform well when there are
a lot of local minima that have the same amplitude and the same width.
Thus it cannot state that the meta-stable states problem has been solved satisfactorily by QA.
However, QA is effective to remove high-lying meta-stable states and to bring the system down to
one of the low-lying states that have small amplitude. From the experience of MOM,\cite{dorado09,jomard08}
we estimate the absolute error in QA might be less than 0.1\,eV. The relative energy difference
between systems would be much better and at an order of $2\sigma$ due to error cancelation.
In practice, QA is an effective method to reduce data scattering and to improve the reliability
of the calculated energetics.

In brief, when the system is small and the computer resource is enough so that allowing to repeat the
calculation many times, MOM is a good choice. With this method one can get explicit information
of how the meta-stable states distribute.
On the other hand, QA optimizes ionic and electronic degrees simultaneously. Thus if atomic structure
optimization is desired, QA would outperform MOM, especially for large systems.
All of the following calculations have been treated by QA procedure to improve the reliability
of the results.

\section{FORMATION ENERGY AND INCORPORATION ENERGY}
\label{sec:EF}

\begin{table*}[]
\caption{\label{tab:Struct-E} LSDA+\emph{U} results for structural and
energetic properties of defects in uranium dioxide:
defect traps [the tri-vacancy (tri-V), the octahedral interstitial site (Int.), the uranium site (U), and the
oxygen site (O)] and xenon-trap aggregates [COT-xe---xenon in COT cluster,
Xe(X)---Xe in trap X], respectively.
$\Delta V$ is the defect induced
volume change that averaged to per fluorite cubic cell (over eight cells totally).
E$^{f}$ is the formation energy, E$^{i}$ is the incorporation energy,
and $q$ is the Bader effective charge of xenon.
The data of three oxygen self-defects
are also included for reference, see Ref.[\onlinecite{geng08b}] for details.
}
\begin{ruledtabular}
\begin{tabular}{l c c c c c c c c c} 
          & COT-xe & Xe(Int.) & Xe(U) & Xe(O) &Xe(tri-V)&tri-V& COT-v & COT-o & O$_{i}$ (Int.)\\
\hline
  $\Delta V$(\AA$^{3}$)  &0.79 &3.74&0.97&3.99 &1.40 &0.54 &$-0.14$&$-1.61$& $-0.29$\\
  E$^{f}$(eV)            &$-4.08$ &9.75&12.92&15.06 &5.17 &4.99 &$-7.18$&$-12.41$&$-2.17$ \\
  E$^{i}$(eV)            &3.10 &9.75&3.87 &7.53 &0.18 &&&& \\
  $q$($|e|$)            &0.20 &0.26&0.96 &0.14 &0.09 &&&& \\
\end{tabular}
\end{ruledtabular}
\end{table*}

The formation energy of a defect $D$ that has $n$ excess oxygen atoms is defined as
\begin{equation}
  E^{f}_{D}=E^{coh}_{D}-E^{coh}_{per}-\frac{n}{2}E_{\mathrm{O}_{2}},
\end{equation}
and for a defect with $m$ excess uraniums is
\begin{equation}
  E^{f}_{D}=E^{coh}_{D}-E^{coh}_{per}-mE_{\alpha\mathrm{U}}.
\end{equation}
Here the cohesive energy $E_{D}^{coh}$ of a defective structure is calculated from its total
energy by subtracting the isolated spin-polarized atomic contributions, and
$E_{per}^{coh}$ is the cohesive energy of the corresponding structure without defect;
$E_{\mathrm{O}_{2}}$ is the binding energy of a neutral dioxygen molecule;
and $E_{\alpha\mathrm{U}}$ is the cohesive energy per atom in the metallic
$\alpha$-U phase.\cite{geng08} For tri-vacancy that keeps the UO$_{2}$ composition unchanged, the formation energy
is given by
\begin{equation}
  E^{f}_{tri}=E^{coh}_{tri}-\frac{N-1}{N}E^{coh}_{per},
\end{equation}
where $N$ is the total number of UO$_{2}$ formula that contained in the perfect cell.
The incorporation energy of xenon is defined by
\begin{equation}
  E^{i}=E^{total}_{Xe(X)}-E^{total}_{X}-E_{Xe},
\end{equation}
where $E^{total}_{Xe(X)}$ is the total energy of a cell in which the xenon atom is at
the trap site $X$, $E^{total}_{X}$ is the total energy of the same cell containing only the trap $X$,
and $E_{Xe}$ is the total energy of an isolated xenon atom.
If taking xenon-trap aggregate as a single defect complex, one can define its formation energy
similar to Eq.(1) or (2) except that here $E_{Xe}$ also should be deducted. Numerically it is equal to
the sum of the trap formation energy and the incorporation energy of xenon at that trap.

The calculated results of structure and energetics of various trap sites and xenon-trap aggregates
in UO$_{2}$ are listed in table \ref{tab:Struct-E}.
The tri-vacancy (tri-V) is a kind of bound Schottky defect that has the same geometry as shown in figure 3(e)
of Ref.[\onlinecite{grimes91}], \emph{i.e.}, a pair of oxygen vacancies that binding with one of its nearest uranium vacancies.
It is believed that this geometry has the lowest energy.\cite{grimes91,nerikar09} Defects $V$-4O$^{''}$, $V$-3O$^{''}$ and ($V$-3O$^{''}$)$_{2}$
are not included here.\cite{geng08c} The trap in these clusters is too small to accommodate
xenon atoms: introducing one xenon atom completely destroys the trap geometry
and the aggregate becomes unstable. In addition, previous investigations showed that these defects have insignificant
concentration comparing with others.\cite{geng08b,geng08c}

The energetic information in table \ref{tab:Struct-E} is interpreted as for isolated
defects. This is appropriate for point defects since the simulation cell is large enough for them.
But for extensive clusters such as COT, the cell is not large enough and the interactions with their images that arising from the periodic
boundary conditions might be remarkable, with dipole-dipole interactions as the leading contribution.
Therefore the obtained energy is more close to the value
of an ordered configuration of the clusters at the
corresponding concentration. Using this energy to describe diluted clusters
is theoretically questionable. But the quality will become better and better as the concentration gets increasing.
Fortunately, COT clusters only appear in the hyper-stoichiometric regime and have high enough concentrations,\cite{geng08b}
which implies that the size-effects of COT clusters might have little impact on our discussion here.

COT-xe has the lowest formation energy (as defined above) in all of the defects considered here,
followed by Xe(tri-V), Xe(Int.),
Xe(U), and Xe(O). The high formation energy of Xe(U) is due to the contribution of
the uranium vacancy trap,
whereas the low value of COT-xe is because of the excess oxygens,
with each one contributing about $-2$\,eV.
Thus it is helpful to divide the xenon-trap aggregate formation energy into two parts: the trap
formation energy and the energy that is required to incorporate xenon atom into the pre-existing trap.
The latter is called incorporation energy and is
also listed in table \ref{tab:Struct-E}.
We see that the most easy trap for xenon to incorporate with is tri-vacancy, followed by
COT cluster and uranium vacancy.

It is understandable that rare gases such as xenon and krypton need a big \emph{space} for them to
be accommodated in the fuel matrix, and gas-fuel incorporation is usually accompanied with drastic swelling
of the latter, especially when the gas atoms occupy mainly the octahedral sites [Xe(Int.)] or oxygen vacancies [Xe(O)].
But if most of the gas atoms go into pre-existing traps that have comparable size,
for example uranium vacancies, tri-vacancies or COT clusters,
the resulting fuel deformation will be much small, as can be perceived from the $\Delta V$ row of table \ref{tab:Struct-E}.
This primary analysis indicates that xenon prefers to COT clusters or uranium vacancies instead
of octahedral sites or oxygen vacancies.
From table \ref{tab:Struct-E} we can see that
the incorporation energy is as high as 9.75\,eV (7.53\,eV) when xenon is at an octahedral site (oxygen vacancy),
and decreases to 3.1\,eV when goes into COT cluster. This suggests that there is
a deep local minimum on the energy surface at COT center,
which will drive xenon atoms from octahedral sites and oxygen vacancies into COT clusters.
Similar conclusion holds for tri-vacancy and uranium vacancy.

\begin{figure}
  \includegraphics*[0.22in,0.19in][4.02in,2.92in]{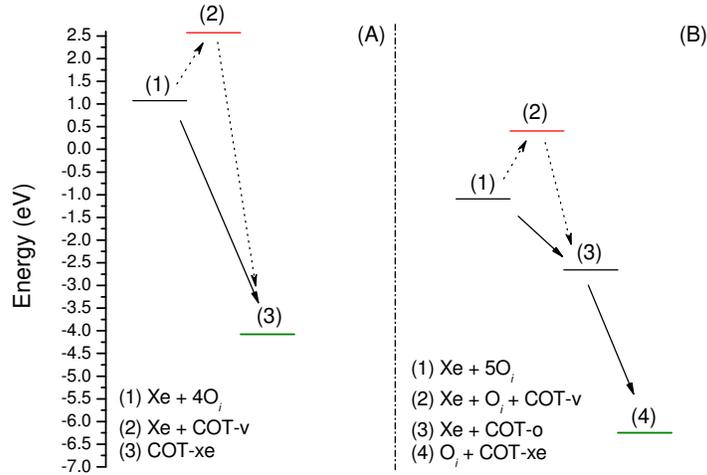}
  \caption{(Color online) Formation energy
  of various defect arrangements in UO$_{2}$:
  (A) a system with one xenon atom and four excess oxygens, and (B) with
  one more excess oxygen. }
  \label{fig:deft-lev}
\end{figure}

Figure \ref{fig:deft-lev} illustrates in details how this takes place in UO$_{2}$,
where the energy variation for different combinations of xenon atom
and excess oxygens is given. For clarity we assumed here that point xenon occupies an octahedral
site. The discussion is similar if it is at an oxygen vacancy.
Point oxygen interstitials are also assumed occupying octahedral sites.
In figure \ref{fig:deft-lev}(B) the
competition between COT-v and COT-o clusters is demonstrated.
The energy cost is about 1.5\,eV to bring four point oxygen interstitials
together to form a COT-v cluster.
The energy gain is 6.5\,eV when the xenon atom
goes into COT-v center from an octahedral site.
Most of this part of energy gain is from the \emph{elastic} contribution.
Similarly, swapping the central oxygen of COT-o with a xenon atom
reduces the total energy by 3.5\,eV.
The preference of xenon to COT trap
is thus evident in both cases.
It is easy to understand this by size
effects: the atomic size of xenon is larger than the octahedral site,
thus a drastic lattice distortion occurs when xenon atom occupying an octahedral site.
But it is not when xenon is in COT clusters.
Therefore as long as there are COT-o or COT-v clusters,
xenon atoms will combine with them instead of occupying octahedral sites or
oxygen vacancies.

Note that there are two kinds of incorporation process for COT-xe: (a) a direct combination of
a xenon atom and a pre-existing COT-v cluster; (b) swap a xenon atom with the central
oxygen of a COT-o cluster. In the first case
the total xenon-trap aggregate formation energy is $-4.08$\,eV, with a xenon incorporation energy of 3.10\,eV,
and a trap formation energy (COT-v) of $-7.18$\,eV, as listed in table \ref{tab:Struct-E}.
In the latter case, however, the
total formation energy is $-6.25$\,eV (with $-2.17$\,eV contributed by the point oxygen
interstitial), with the xenon incorporation energy of
6.16\,eV, and the trap formation energy (COT-o) of $-12.41$\,eV.
Therefore in the case (a) the incorporation is easy but the available number of
COT-v trap is rare, whereas in the case (b) the situation is just opposite.

\begin{figure}
  \includegraphics*[width=3.0 in]{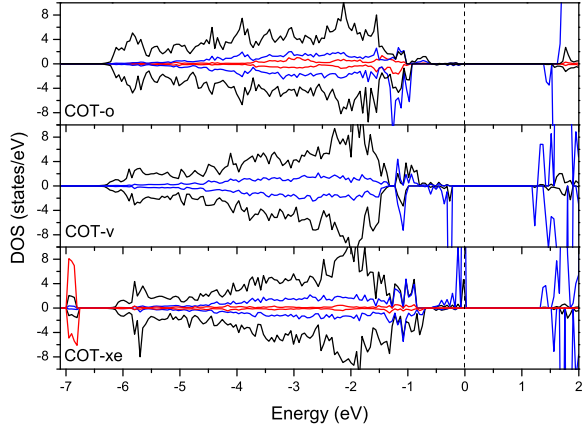}
  \caption{(Color online) Density of states of O 2$p$ (black) and U 5$f$ (blue) orbitals
  that projected onto the defect atoms in COT-o (upper panel), COT-v (middle panel), and COT-xe (lower panel), respectively.
  Contribution from the central atom (oxygen in COT-o and xenon in COT-xe) is marked by red lines.
  The dotted vertical line indicates the Fermi level.}
  \label{fig:deft-dos}
\end{figure}

To understand this kind of incorporation behavior,
the mechanism of the energy variation from COT-o to COT-v is the key.
Figure \ref{fig:deft-dos} shows the electronic density
of state (DOS) of oxygen 2$p$ and uranium 5$f$ orbitals that projected onto
the cuboctahedron atoms of COT-o, COT-v and COT-xe, respectively.
The red lines indicate the contribution from the oxygen or xenon atom that occupies the
cluster center, and the dotted vertical line indicates the Fermi level.
There are three significant features:
(1) the distribution profile of the O 2$p$ DOS within the main valence band, (2) the localized U 5$f$
states just below the Fermi level, and (3) the interplay of the localized xenon state and the valence band.

In COT-o, the central oxygen not only hybridizes itself with U 5$f$ states directly, but also
enhances the overlapping of the main O 2$p$ and U 5$f$ orbitals near the upper band edge.
As a result, the localized U 5$f$ states just below the Fermi level are absorbed
into the valence band. This makes COT-o stable in energetics. In contrast, in COT-v and COT-xe, the hybridization
with localized U 5$f$ states is insufficient. Though a weak anti-bond is formed
by partial overlapping, a highly localized state still presents near the Fermi level.
This feature and the fact that O 2$p$ DOS distributes more on the upper part in the main valence band
reveal the electronic origination of the energy increase from COT-o to COT-v.
Different from oxygen whose effect is at the upper band edge,
xenon in COT-xe affects mainly the lower edge
of the valence band. As shown in the lower panel of figure \ref{fig:deft-dos}, xenon
interacts with itinerant O 2$p$ and U 5$f$ orbitals weakly, splitting
the main valence band into a minor localized bonding state and a major anti-bonding band that is dispersive.
The latter is then pushed to higher energy and is the main
contribution to the energy difference between COT-xe and COT-v.
Since this band shift keeps most features of the valence band unchanged, xenon effects
on energetics is thus mainly mechanical.

Table \ref{tab:Struct-E} also lists the Bader effective charge of xenon in
various traps at the last row.\cite{henkel06} We see that except tri-vacancy in which xenon is charge neutral,
all other traps have a tendency to \emph{ionize} xenon atom.
Usually the degree of this ionization is small
when the gas atom is in COT cluster,
oxygen vacancy, or octahedral site.
This is because of the relatively electron-rich environment of these traps.
But in the case of Xe(U), one electron has been completely peeled off
due to the strong Madelung potential of the periodic lattice at the uranium vacancy sites.
It is a competition process between the
electron affinity of the trap and the ionization energy of the gas atom.
Thus the ionization does not lead to a large energy increase.
The observed ionization of Xe to Xe$^{+}$ confirms
Grimes \emph{et al.}'s prediction with shell model.\cite{grimes91}
This charge transfer has important implication on gas bubble growth rate.
It is believed that gas bubble grows by accumulation of point xenon impurities via diffusion.
Positively charged gas atoms expel each other strongly due to the electrostatic interactions.
Thus they cannot approach together.
This means that inert gas bubble cannot initiate from uranium vacancies.
In agreement with shell model results, we found that fission gas bubbles can only start
from neutral tri-vacancy sites.\cite{jackson85} As will be discussed below,
this property makes temperature and the chemical composition
being effective parameters to tune the growth rate of bubbles.

\section{SOLUTION ENERGY AT FINITE TEMPERATURES}
\label{sec:ES}

\begin{figure}
  \includegraphics*[width=3.0 in]{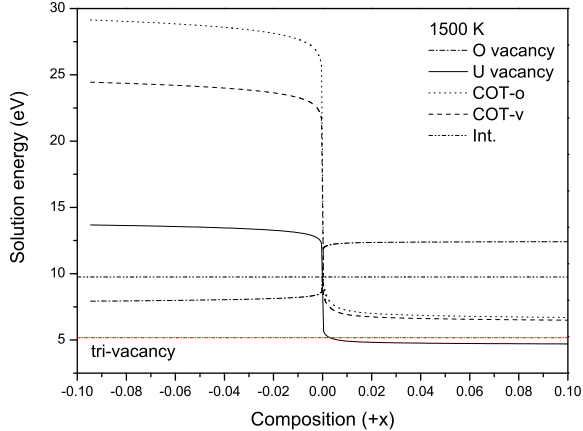}
  \caption{Solution energy of xenon at various traps in UO$_{2}$ at 1500 Kelvin
  }
  \label{fig:deft-se}
\end{figure}

With only the information of formation energy and incorporation energy
that discussed in the previous section,
it is difficult to evaluate what kind of trap the gas atoms prefer to at finite temperatures.
For example COT-xe has the lowest trap formation energy---which suggests a high concentration of the
trap---but the incorporation energy is high.
Thus it is not necessarily the favored one,
because the preferred site should be a combination result of the available number of the trap sites
and the degree of incorporation difficulty.
We knew from previous investigations on oxygen clustering behavior in UO$_{2}$
that in hypo-stoichiometric regime (UO$_{2-x}$) oxygen vacancy is the major
defect,\cite{geng08} and in UO$_{2+x}$ there is a transition from point oxygen interstitials
to COT-o clusters with an increase of the composition.\cite{geng08b,geng08c}
That is, oxygen vacancy and COT-o are the most available traps in UO$_{2}$.
On the other hand, the concentrations of COT-v, uranium vacancy and tri-vacancy are
one order smaller, but they have lower incorporation energies. Therefore
a delicate analysis is required in order to get the final answer.

Theoretically, the probability for a
gas atom to be trapped in a specific site is proportional to the product of the trap
concentration $\rho_{t}$ and the probability to incorporate the gas atom
into one of these traps.
In this way one can define the \emph{solution}
\emph{energy} as
\begin{equation}
\label{eq:Es}
E^{s}=-\kappa_{B}T\ln \rho_{t}+E^{i},
\end{equation}
which gives the probability of trapping one gas atom to a specific trap
by $\exp(-E^{s}/\kappa_{B}T)$.
Here $\kappa_{B}$ is the Boltzmann constant and $T$ is the temperature. The first term
in Eq.(\ref{eq:Es}) gives the effective formation free energy of the trap
and the second term is the incorporation energy.
With independent cluster approximation,\cite{geng08,geng08b,geng08c,matzke87,lidiard66}
one can evaluate the trap concentration
as a function of temperature and chemical composition in the closed regime.
This approximation holds valid as long as there has no explicit \emph{overlapping}
or strong interactions
among traps.
Note that the temperature effect considered here is only the statistical effect on trap concentrations.
Contribution from lattice vibrations on
defect formation energy can be included straightforwardly, but here we did not take this into account because we
focus mainly on the fundamental incorporation behavior of single gas atoms.

\begin{figure}
  \includegraphics*[width=3.0 in]{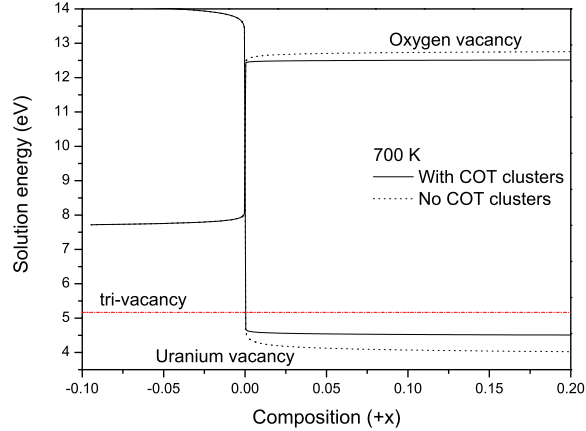}
  \caption{Influence of oxygen defect clustering on solution energy of xenon at 700 Kelvin.
  }
  \label{fig:deft-se2}
\end{figure}

Calculated solution energies of xenon at various trap sites
at 1500\,K are shown in figure \ref{fig:deft-se}.
For tri-vacancy and octahedral interstitial sites, the solution energies are independent of composition
since creation of these traps does not change the stoichiometry.
The stoichiometric effects on oxygen vacancy and uranium vacancy are
opposite. There is a step-like jump near the stoichiometry. Far away from that composition,
the variation of solution energy flats out.
In hypo-stoichiometric regime, tri-vacancy has the lowest solution energy.
All other traps have a solution energy at least 3\,eV higher,
and the physical picture is clear and trivial. In the other side of the stoichiometry, however, thermodynamic competition
becomes important.
The solution energy of xenon at tri-vacancy sites is 5.15\,eV.
The solution energy for xenon at uranium vacancy sites is about 4.7\,eV,
slightly lower than that at tri-vacancy. This is different from shell model
results which predicted that cation vacancy had a much lower solution energy than tri-vacancy.\cite{grimes91}
With LSDA+\emph{U} calculations, though uranium vacancy has the lowest solution energy,
the tri-vacancy is at a level of \emph{just} 0.4\,eV higher.
In contrast, COT clusters are about 1.8\,eV higher in solution energy, mainly due to the few concentration of COT-v
trap in the incorporation process (a) and the great incorporation energy
in the case (b).

When decrease the temperature to 700\,K, because of the change in trap concentrations, the
solution energy of xenon at uranium vacancy sites reduces 0.7\,eV, as shown in figure \ref{fig:deft-se2}.
This change is important since it almost excludes the possibility for xenon atoms to occupy
tri-vacancy sites.
The influence of oxygen clusters is also non-trivial here. At 700\,K, inclusion of COT clusters increases
the solution energy of xenon at uranium vacancy sites by a value of 0.5\,eV to 4.5\,eV. This effect takes 43\% of the solution energy
difference between xenon at uranium vacancy and at tri-vacancy sites, thus is significant.\cite{cluster}
Furthermore, since concentration of oxygen defect clusters are inversely proportional to temperature,\cite{geng08b,geng08c}
at lower temperatures
this indirect influence will become more distinct.

For all trap sites that were investigated in this work, the solution energies are positive.
It indicates the insolubility of xenon gas in bulk UO$_{2}$.
But as a fission reaction product, xenon keeps in the fuel matrix until it diffuses to grain boundaries and forms
large bubbles. Before that, gas atoms might distribute randomly in the material and grow into small bubbles at defect sites
or on dislocation loops.
The moderate solution energy difference of xenon between at uranium vacancy and at tri-vacancy sites in $x>0$ regime
suggests that one can change temperature to \emph{tune} the incorporation behavior of xenon, and then the bubble growth
rate.\cite{note,sattonnay06} This provides an alternative point to investigate the fuel swelling mechanism and the consequent structure damage
imposed to the fuel and/or the cladding.

\section{COMPARISON WITH OTHER THEORETICAL CALCULATIONS}
\label{sec:comp}

\begin{table}[]
\caption{\label{tab:inc-E} Calculated incorporation energies of xenon in UO$_{2}$ at
various trap sites by different methods: the uranium site (U), the oxygen site (O), the
octahedral interstitial site (Int), the tri-vacancy site (tri-V) and the cuboctahedron site (COT).
}
\begin{ruledtabular}
\begin{tabular}{l c c c c c } 
   $E^{i}$ (eV)& COT-xe & Xe(Int.) & Xe(U) & Xe(O) & Xe(tri-V)  \\
\hline
  LSDA+\emph{U}\footnotemark[1] &3.10 & 9.75&3.87 & 7.53& 0.18 \\
  GGA+\emph{U}\cite{nerikar09}  &$-$ & 11.11&2.5 & 9.5& 1.38 \\
  GGA+\emph{U}\cite{yu09}    &$-$ & 8.07 &5.18&9.01&2.90  \\
  GGA\cite{nerikar09}    &$-$ & 12.75&6.04&9.71&2.12   \\
  GGA\cite{freyss06}    &$-$ & 11.2 &13.9& 9.4&$-$  \\
  Shell model\cite{grimes91} &$-$ & 17.23&4.99&13.34 &1.16  \\
  Shell model\cite{nicoll95} &$-$ & 18.67&5.83&15.15 &3.3  \\
\end{tabular}
\footnotetext[1]{\, this work}\\
\end{ruledtabular}
\end{table}

There were several DFT calculations on the incorporation behavior of xenon in UO$_{2}$ published recently.\cite{freyss06,nerikar09,yu09}
It is valuable to summarize these results and compare with our calculations.
The results of DFT calculations and those computed using semi-empirical shell model potentials\cite{grimes91,nicoll95}
are listed in table \ref{tab:inc-E}.
These data are scattered, but a common trend is obvious. All calculations that conducted at different
theoretical levels predicted that the lowest incorporation energy for xenon is at the tri-vacancy site,
and the highest energy is at the octahedral interstitial site, which is then followed by the oxygen and the uranium vacancy sites.
Only the GGA results of Ref.[\onlinecite{freyss06}] predicted a different order. Its Xe(U) has
an incorporation energy as high as 13.9\,eV. This might originate from size effect since they employed a small
supercell containing only 12 atoms. In a recent GGA calculation using a larger supercell containing 96 atoms,\cite{nerikar09}
this value was reduced to 6.04\,eV and became qualitatively consistent with other calculations.
In addition, there is a digital scattering with a magnitude of 1$\sim$2\,eV in these DFT results. This might be due to the different
methodologies that were employed, for example the different exchange-correlation functionals, the constraints on structure optimization
process, the effective Hubbard parameters, and so on. Meta-stable states might also play some
role here.

The incorporation energies of xenon calculated by Nerikar \emph{et al.}\cite{nerikar09}
are 1$\sim$2\,eV higher than ours, except that at uranium vacancy where their value is about 1\,eV
smaller. As for the solution energies, they predicted the stability of xenon at tri-vacancy in the
hypo-stoichiometric regime with a solution energy of 3.88\,eV, about 2\,eV smaller than ours.
This is because though we have a similar binding energy for point defects to form a tri-vacancy,
our formation energies of individual point defects are higher than theirs. In hyper-stoichiometric
regime, they and we both predicted that uranium vacancy is the favored site for xenon at low temperatures.
Again our value is about 2\,eV higher. Without considering oxygen clustering and finite temperature
effects, Nerikar \emph{et al.} failed to notice the variation of the solution energy with temperature
and the thermodynamic competition among tri-vacancy, uranium vacancy and other possible
complex traps (\emph{e.g.}, di-vacancy), which we have shown are important for understanding the physical behavior of
realistic nuclear fuels.

\section{CONCLUSION}

With LSDA+\emph{U} calculations, the incorporation behavior of xenon atoms at various trap sites
in UO$_{2}$ was analyzed by studying the electronic structure and
energetics. In the regime of UO$_{2-x}$, the result was that the gas atom prefers to
tri-vacancy sites, while in UO$_{2+x}$
uranium vacancies are favored.
COT clusters have large enough space at their centers but are not occupied by xenon atom due to subtle electronic
de-hybridization of O $2p$ and U $5f$ orbitals when xenon atom goes in and the central oxygen out.
The calculated solution energies showed that a
thermodynamic competition between Xe(U) and Xe(tri-V) is significant,
and the indirect influence of oxygen clustering is important.
At uranium vacancy site, xenon is ionized to Xe$^{+}$ state, confirmed early semi-empirical
prediction. This kind of charge transfer indicates that xenon bubbles cannot initiate from uranium vacancies but neutral
tri-vacancy sites. Thus one can tune the occupation probability of xenon at tri-vacancy sites and
then the bubble growth rate by control
temperature.
A approach to solve the meta-stable states difficulty that frequently encountered in DFT+\emph{U} applications
was proposed, which exploits the coupling between ionic and electronic sub-systems and uses a quasi-annealing procedure
to relax the electronic system to the ground state. It was shown that this method can effectively avoid meta-stable states.

\begin{acknowledgments}
Support from the Fund of National Key Laboratory of Shock Wave and Detonation Physics of China
(under Grant No. 9140C6703031004) is acknowledged.
The work was also partially supported by the Budget for
Nuclear Research of the Ministry of Education, Culture, Sports,
Science and Technology of Japan (MEXT), based on the screening and counseling by the
Atomic Energy Commission, and by the Next Generation
Supercomputing Project, Nano-science Program, MEXT, Japan.
\end{acknowledgments}



\begin{thebibliography}{99}
\bibitem{note} Temperature also affects the diffusivity of defects and micro gas bubbles, which in turn determines the
development rate of grain boundary bubbles.
\end{thebibliography}
\end{document}